How to improve the prediction based on citation impact percentiles for years shortly after the publication date?

Lutz Bornmann,[#] Loet Leydesdorff,[*] and Jian Wang[§+]

[#] Division for Science and Innovation Studies

Administrative Headquarters of the Max Planck Society

Hofgartenstr. 8,

80539 Munich, Germany.

E-mail: bornmann@gv.mpg.de

[*] Amsterdam School of Communication Research (ASCoR),

University of Amsterdam, Kloveniersburgwal 48,

1012 CX Amsterdam, The Netherlands; loet@leydesdorff.net

[§] Institute for Research Information and Quality Assurance (iFQ)

Schützenstraße 6a, 10117 Berlin, Germany

[+] Center for R&D Monitoring (ECOOM) and Department of Managerial Economics, Strategy

and Innovation, Katholieke Universiteit Leuven,

Waaistraat 6, 3000 Leuven, Belgium; Email: Jian.Wang@kuleuven.be

**Abstract**

The findings of Bornmann, Leydesdorff, and Wang (in press) revealed that the consideration of journal impact improves the prediction of long-term citation impact. This paper further explores the possibility of improving citation impact measurements on the base of a short citation window by the consideration of journal impact and other variables, such as the number of authors, the number of cited references, and the number of pages. The dataset contains 475,391 journal papers published in 1980 and indexed in Web of Science (WoS, Thomson Reuters), and all annual citation counts (from 1980 to 2010) for these papers. As an indicator of citation impact, we used percentiles of citations calculated using the approach of Hazen (1914). Our results show that citation impact measurement can really be improved: If factors generally influencing citation impact are considered in the statistical analysis, the explained variance in the long-term citation impact can be much increased. However, this increase is only visible when using the years shortly after publication but not when using later years.

# 1 Introduction

Percentiles used in bibliometrics provide information about the citation impact of a focal paper compared with other comparable papers in a reference set; for example, all papers in the same research field and publication year. A percentile is the value below which a certain proportion of observations (here: papers) fall: the larger a paper's percentile, the higher citation impact it has – compared with papers in the same field and publication year. Since the percentile approach has been acknowledged in bibliometrics as a valuable alternative to the normalization of citation counts based on mean citation rates, some different percentile-based approaches have been developed (see an overview in Bornmann, Leydesdorff, & Mutz, 2013). More recently, two of these approaches ($PP_{top\ 10\%}$ and the Excellence Rate, respectively) have been prominently used in the Leiden Ranking (Waltman et al., 2012) and the SCImago institutions ranking (SCImago Reseach Group, 2012) as evaluation tools.

Using a publication set including all papers published in 1980 (nearly 500,000 papers), Bornmann, et al. (in press) investigated how the different percentile-based approaches are able to predict the long-term citation impact (in year 31, $t_{31}$) of papers from citation impacts in previous years (years 1, $t_1$, to 30, $t_{30}$). In comparison to the other approaches, the SCImago approach demonstrated unexpected capabilities in accurately predicting the long-term citation impact on the basis of the citation impact in the first few years after publication. The consideration of the journal impact in this approach in solving the problem of tied citations seems to have generated this positive effect.

For the problem of ranks tying at the top 10% threshold level (Bornmann, de Moya Anegón, & Leydesdorff, 2012), SCImago introduces a secondary sort key in addition to citation counts: When citation counts are equal, the publication in a journal with the higher SCImago Journal Rank (SJR2) (Guerrero-Bote & de Moya-Anegon, 2012) obtains the higher percentile rank. Adding this journal metric takes into account not only the observed citations of the focal paper but also the prestige of the journal that a paper is published in.

Given the enduring tension between the practical needs for timely assessment of research outputs and the long time period it takes for research to reveal its full impact (Bornmann, 2013; Wang, 2013), we examine in the present study the added-value of considering the journal impact in predicting the long-term citation impact from citation impacts in previous years. For the statistical analyses, we use the same data set as Wang (2013) and Bornmann, et al. (in press). However, we consider not only the journal impact but also other factors (e.g., the number of authors) for better predicting long-term citation impact. Bibliometric studies have already pointed out several other factors –in addition to journal impact − with an (significant) effect on citation impacts (see an overview in Bornmann & Daniel, 2008). Thus, we examined, whether the prediction of long-term citation impact (based on years shortly after the publication date) can be improved by considering further factors. Since the approach of Hazen (1914) to calculate percentiles is widely used in statistical packages, we use it in this study. The results are also generalizable to other approaches (as we will exemplarily show).

## 2 Methods

### 2.1 The percentile approach of Hazen (1914)

Two steps are needed in order to calculate percentiles for a reference set based on the percentile-based approach of Hazen (1914):

First, all papers in the set are ranked in ascending order of their numbers of citations. Papers with equal citation counts are set equal by assigning the average rank. This is the default ranking method in the statistical package Stata (StataCorp., 2011). This method ensures that the sum of the ranks is fixed at $n*(n+1)/2$, where $n$ is the number of papers in the reference set.

Second, each paper is assigned a percentile based on its rank (percentile rank). Percentiles can be calculated in different ways (Bornmann, Leydesdorff, et al., 2013; Cox, 2005; Hyndman & Fan, 1996). In this study, we used the formula (100 * ($i$-0.5)/$n$), derived by Hazen (1914). This formula is used very frequently nowadays for the calculation of percentiles and is wired into the official Stata command "quantile" (StataCorp., 2011). It ensures that the mean percentile is 50 and symmetrically handles the tails of the distributions.

**2.2   Dataset used**

In this study, we define a reference set for a paper under study as a set of papers with the same WoS subject category and document type. The reference sets were used to calculate the percentile-based approach developed by Hazen (1914). Each paper in WoS is classified into one unique document type but possibly into multiple subject categories. Therefore, for papers with multiple subject categories, the average percentile rank is used.

Furthermore, the citation percentiles could be too coarse if the size of the reference set is too small. Therefore, only reference sets with at least one hundred papers are included.[1] For example, if a paper belongs to two different reference sets: *A* and *B*, and *A* has more than 100 papers while *B* has less than 100 papers, then the percentiles based on *B* are discarded. If neither *A* nor *B* has more than 99 papers, then both results based on *A* and *B* are discarded, and this paper is excluded from the further analysis.

The dataset contains all journal papers published in 1980 and indexed in Web of Science (WoS, Thomson Reuters), that is, 746,460 papers in total. Two restrictions are then imposed on the sample: (1) three document types – articles, reviews, and notes [2] – were kept while other documents types were excluded, and (2) only papers having at least one reference set with hundred or more papers were included. As a result, we have 475,391 papers for analysis, and the annual citation counts (from 1980 to 2010) for these papers were retrieved from WoS.

---

[1] We decided to use 100 papers as a limit to produce reliable data. There is a high probability that the use of a limit of 50 or 200 would come to similar results as ours.
[2] Notes were removed from the database as a document type in 1997, but they were citable items in 1980.

## 2.3 Statistical procedures and variables (covariates)

We fitted 30 sets of regression models with the percentile of citations in year 31, $t_{31}$, as the dependent variable and the (short-)time-window citation percentiles (from year 1 to year 30) as one independent variable, correspondingly. For each set of models, Model 1 only uses short-time-window citation percentiles as predictors (for example, model 1 in set 1 only uses citation percentile in year 1 as the predictor), and Model 2 to 5 sequentially add the other four covariates: journal impact factor (JIF), number of authors, number of cited references, and number of pages. We used ordinary least-squares regressions, and the normality assumption is not seriously violated. The JIF in the year of 1983 is used and calculated from the database by the authors: $JIF_{1983}$ for a journal equals to the number of times that "citable items" published in that journal in 1981 and 1982 were cited in 1983 divided by the total number of "citable items" published by that journal in 1981 and 1982. The "citable items" include articles, reviews, and notes. Previous studies have shown that an increase in the number of authors, the number of cited references, and the number of pages can be expected to result in a higher impact of a paper (Bornmann & Daniel, 2008; Didegah & Thelwall, 2013; Taborsky, 2009; Vieira & Gomes, 2010). We expect that the consideration of these proven relationships in the regression models will lead to a better prediction of the long-term citation impact – especially based on the citation impact measured shortly after the publication date.

The number of authors, the number of cited references, and the number of pages are skew distributed, so the natural logarithm of these variables is used for model estimations. In addition, using the log also allows non-linear relationships between the dependent and the independent variables. Previous studies suggest diminishing effects or even inverted-U shaped effects of these factors (Bornmann & Williams, 2013). We empirically tested the linear models (using the original scale of these variables), diminishing return models (using the log of these variables), and inverted-U shaped models (using the original scale of these variables and their squared terms) for each variable individually, and found that the log models result in the lowest Bayesian information criterion (BIC) values. The BIC has been proposed as a means to compare the fit of different regression models. The lower the value of BIC, the better the fit (Long & Freese, 2006). This result suggest that the effects of the number of authors, the number of cited references, and the number of pages on citations (ln) are better described by diminishing return models than by linear or quadratic models. In other words, as the number of authors, references, or pages increases, citations (ln) increase at a decreasing rate. Therefore the natural logarithm of these variables is used for model estimations, instead of using the original scale of these variables or using the linear and squared terms of these variables.

## 3    Results

Table 1 reports descriptive statistics of the dependent variables and covariates. Because shorter citation time windows are of greater interest in this context, we only report model set 1 and set 2 here, in Table 2 and Table 3, respectively. As discussed before, from model 1 to model 5, covariates are gradually added. All regression coefficients are statistically significant, and it might be because of the large sample we have (Kline, 2004). With regard to the research question of this study (how to improve the prediction of the long-term citation impact?), the most important information is the adjusted $R^2$ in Table 2, which is an indicator for the explanatory power of a model and "represents the squared residuals that are explained by the model as a share of the total squared residuals" (Kohler & Kreuter, 2012, p. 269). We calculated an adjusted $R^2$, because $R^2$ monotonically increases as terms are added to a model (Hardin & Hilbe, 2012). An adjusted $R^2$ includes shrinkage terms. The higher the (adjusted) $R^2$ in Table 2, the more variation in citation impact at time $t_{31}$ is explained by the covariates, including the citation impact at time $t_1$ and other factors found to have considerable effects on citation impact (e.g., the number of authors) in the literature.

As the results in Table 2 show, adding JIF increases the $R^2$ from .09 (model 1) to .19 (model 2), while further adding the number of authors (ln) does not have additional effect on the $R^2$. However, the number of cited references (ln) and the number of pages (ln) do further improve the $R^2$. If all factors (covariates) are included in the regression model (model 5), the $R^2$ reaches the highest value of .25. This result shows that incorporating these factors leads to a significantly improved prediction of the long-term citation impact at a very early time point: the first year after the publication of a paper.

Table 3 shows the regression model sets predicting citation impact in year 31 based on citation impact in year 2 gradually adding covariates. In comparison to Table 2, the citation impact at $t_2$ can already explain a relatively high amount of the long-term citation impact: the $R^2$ is .35. Adding the covariates having a general influence on citation impact leads to an increase in the $R^2$ up to .43. Thus, the consideration of these covariates can improve the $R^2$ but not to the extent at year 1 (see the results for year 1 in Table 2).

Figure 1 shows the variance ($R^2$) of citation impact at time $t_{31}$ which is explained by the covariates in the different models (citation impact at time $t_1$ to $t_{30}$ as well as influencing factors of citation impact). Corresponding to each model in Table 2 and Table 3, 28 further models were calculated using the citation impact at time $t_3$ to time $t_{30}$ (instead of the citation impact at time $t_1$ and $t_2$) as a covariate. It is clearly visible in Figure 1 that the estimation of the long-term citation impact can only be improved by the consideration of the covariates in the years shortly after the publication date (the first one to three years). In later years, inclusion of the covariates no longer improves the $R^2$ of the long-term citation impact.

We repeated the calculation of the regression models using a citation rank approach (P100) (Bornmann, et al., in press) instead of the percentiles based on the formula of Hazen (1914) and received very similar results (which are therefore not shown in this paper).

# 4    Discussion

The findings of Bornmann, et al. (in press) revealed that the consideration of journal impact improves the prediction of long-term citation impact. In this study we have tried to investigate whether it is possible to improve citation impact measurements on the base of a short citation window by the consideration of journal impact and other factors, such as the number of authors, the number of cited references, and the number of pages. Our results show that citation impact measurement can really be improved in this manner in the first years after publication. If factors generally influencing citation impact are considered in the statistical analysis, the variance in the long-term citation impact explained by the citation impact in years shortly after the publication date of a paper can be much increased. However, this increase is only significant when using the years shortly after publication, but disappears in later years.

Our results suggest that in bibliometric evaluation that use short citation time windows, citation impact measurements should be adjusted for factors influencing citation impact. Similar adjustments have been proposed for institutional performance measurements: Starting out from Goldstein and Spiegelhalter's (1996) recommendations for the conducting of quantitative comparisons among institutions, Bornmann, Mutz, and Daniel (2013) undertook a reformulation of the Leiden Ranking 2011/2012 (LR) by means of multilevel regression models. The LR was published early in 2012 and uses a set of bibliometric indicators to rank the universities. The results of the regression models show that the larger the publication output of a university and the greater the number of inhabitants, total area, and GDP (PPP) per capita of a country where the university is located, the higher the citation impact of a university. This result is comparable to the findings in this study where we found that the prediction of the long-term citation impact of a paper is significantly influenced by several factors (e.g. the JIF).

Based on the results of the regression models, Bornmann, Mutz, et al. (2013) generated a <u>covariate-adjusted</u> ranking of universities. This ranking shows the differences among the universities assuming that all universities have the same mean in each of the covariates included. The results of the covariate-adjusted ranking differ greatly from the non-adjusted ranking. The results of a correlation analysis also make this clear: There is only a moderate correlation (r = .47) between the results of the non-adjusted and covariance-adjusted performance metrics. Similar to the procedure of Bornmann, Mutz, et al. (2013) for universities, covariance-adjusted citation impact metrics could be generated for single papers, which could then be used instead of the non-adjusted impact metrics. As our results point out the adjusted metrics would be a better predictor of the long-term citation impact of a publication than the non-adjusted metrics.

It could be the task of future research to generate these adjusted citation impact metrics and to validate them against an external criterion. For example, Bornmann and Leydesdorff (2013) used data of F1000 to investigate the relationship between peers' ratings and bibliometric metrics. F1000 is a post-publication peer review system of the biomedical literature. The comparison of metrics with peer evaluation has been widely acknowledged as a way of validating metrics:[3] "The natural criterion against which to validate metrics is expert evaluation by peers" (Harnad, 2008, p. 103). A successful validation of the adjusted citation impact metrics in addition to the results referring to reliability in this study would confirm the necessity to consider influencing factors on citation impact in bibliometric studies (especially in years shortly after the publication date of a paper).

---

[3] When using peer evaluation to validate metrics, it should be considered that peers are also fallible (Bornmann, 2011)

# Acknowledgements

The data used in this paper are from a bibliometrics database developed and maintained by the Competence Center for Bibliometrics for the German Science System (KB) and derived from the 1980 to 2011 Science Citation Index Expanded (SCI-E), Social Sciences Citation Index (SSCI), Arts and Humanities Citation Index (AHCI), Conference Proceedings Citation Index- Science (CPCI-S), and Conference Proceedings Citation Index- Social Science & Humanities (CPCI-SSH) prepared by Thomson Reuters (Scientific) Inc. (TR®), Philadelphia, Pennsylvania, USA: ©Copyright Thomson Reuters (Scientific) 2012. KB is funded by the German Federal Ministry of Education and Research (BMBF, project number: 01PQ08004A).

Table 1.

Description of the dependent variable and covariates (*n*=475,391 publications)

| Variable | Mean | Standard deviation | Minimum | Maximum |
|---|---|---|---|---|
| Percentiles in year 31 | 49.68 | 27.95 | 0.70 | 100.00 |
| Percentiles in year 1 | 50.02 | 16.25 | 33.45 | 100.00 |
| Percentiles in year 2 | 49.97 | 24.06 | 7.55 | 100.00 |
| JIF | 2.08 | 1.52 | 0.00 | 26.98 |
| Number of authors (ln) | 0.69 | 0.59 | 0.00 | 4.57 |
| Number of cited references (ln) | 2.35 | 1.19 | 0.00 | 7.24 |
| Page numbers (ln) | 1.94 | 0.67 | 0.00 | 4.62 |

Table 2.

Regression models predicting citation impact at year 31 based on citation impact <u>in year 1</u> gradually adding covariates

|  | Model 1 | Model 2 | Model 3 | Model 4 | Model 5 |
|---|---|---|---|---|---|
| Intercept | 24.36 * (0.13) | 17.15 * (0.12) | 15.72 * (0.13) | 7.14 * (0.13) | 1.68 * (0.15) |
| Percentile in year 1 | 0.51 * (0.00) | 0.40 * (0.00) | 0.40 * (0.00) | 0.38 * (0.00) | 0.37 * (0.00) |
| Journal Impact Factor |  | 5.95 * (0.02) | 5.64 * (0.03) | 4.17 * (0.03) | 4.35 * (0.03) |
| Number of authors (ln) |  |  | 3.06 * (0.06) | 1.85 * (0.06) | 2.20 * (0.06) |
| Number of cited references (ln) |  |  |  | 5.82 * (0.03) | 4.73 * (0.04) |
| Page numbers (ln) |  |  |  |  | 3.95 * (0.06) |
| $R^2$ (adjusted) | 0.09 | 0.19 | 0.19 | 0.24 | 0.25 |

Notes.
Standard errors in parentheses
* $p<.01$

Table 3.

Regression models predicting citation impact at year 31 based on citation impact <u>in year 2</u> gradually adding covariates

|  | Model 1 | Model 2 | Model 3 | Model 4 | Model 5 |
|---|---|---|---|---|---|
| Intercept | 15.17 * (0.08) | 11.27 * (0.08) | 10.23 * (0.08) | 3.92 * (0.09) | -0.50 * (0.11) |
| Percentile in year 2 | 0.69 * (0.00) | 0.62 * (0.00) | 0.61 * (0.00) | 0.59 * (0.00) | 0.58 * (0.00) |
| Journal Impact Factor |  | 3.64 * (0.02) | 3.41 * (0.02) | 2.39 * (0.02) | 2.54 * (0.02) |
| Number of authors (ln) |  |  | 2.33 * (0.06) | 1.44 * (0.05) | 1.72 * (0.05) |
| Number of cited references (ln) |  |  |  | 4.48 * (0.03) | 3.61 * (0.03) |
| Page numbers (ln) |  |  |  |  | 3.18 * (0.05) |
| $R^2$ (adjusted) | 0.35 | 0.39 | 0.39 | 0.42 | 0.43 |

Notes.
Standard errors in parentheses
* p<.01

Figure 1.

Explained variance of citation impact at year 31 by the covariates in the different models

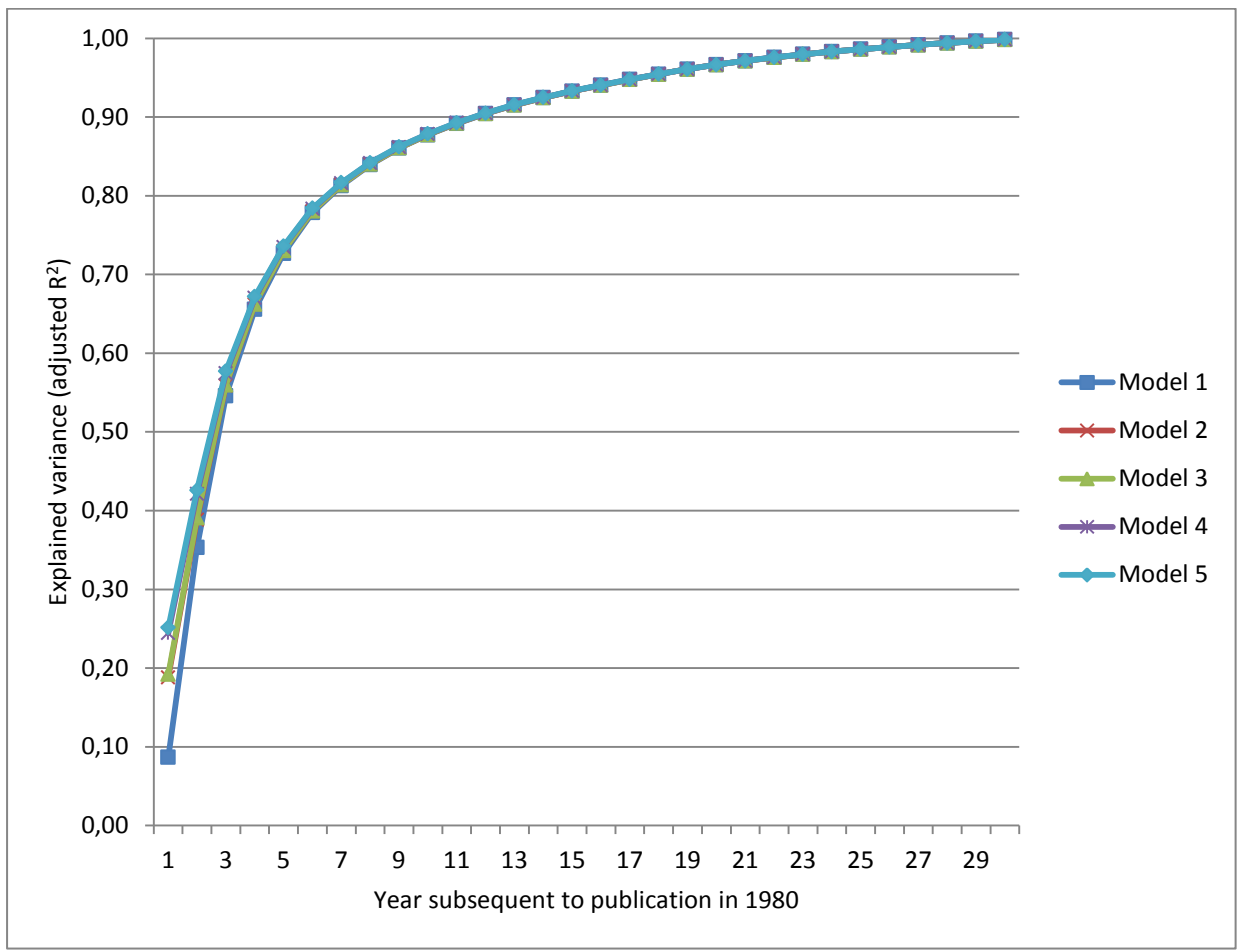